\documentclass[onecolumn,secnumarabic,amssymb, nobibnotes, aps, prd]{revtex4-2}
\usepackage{graphicx}
\usepackage{amsmath}
\usepackage{amsthm}
\usepackage{amsfonts}
\usepackage{amssymb}
\usepackage{mathrsfs}
\usepackage{color}
\usepackage{longtable}

\newcommand{\env}[1]{\texttt{#1}}
\begin{document}

\title{A test of gravitational theories including torsion with the BepiColombo radio science experiment}%

\author{Giulia Schettino}%
\email[Giulia Schettino: ]{g.schettino@ifac.cnr.it}
\affiliation{IFAC - CNR, Via Madonna del Piano 10, 50019 Sesto Fiorentino (FI), Italy, and Dipartimento di Matematica, Universit\`a di Pisa, Largo Bruno Pontecorvo 5, 56127 Pisa, Italy}

\author{Daniele Serra}%
\affiliation{Dipartimento di Matematica, Universit\`a di Pisa, Largo Bruno Pontecorvo 5, 56127 Pisa, Italy}

\author{Giacomo Tommei}%
\affiliation{Dipartimento di Matematica, Universit\`a di Pisa, Largo Bruno Pontecorvo 5, 56127 Pisa, Italy}

\author{Vincenzo Di Pierri}%
\affiliation{Dipartimento di Matematica, Universit\`a di Pisa, Largo Bruno Pontecorvo 5, 56127 Pisa, Italy}

\date{\today}%
\begin{abstract}
  Within the framework of the relativity experiment of the ESA
  BepiColombo mission to Mercury, launched at the end of 2018, we
  describe how a test of alternative theories of gravity including
  torsion can be set up. Following March et al. (2011), the effects of a
  non-vanishing spacetime torsion have been parameterized by 3
  torsion parameters, $t_1$, $t_2$, $t_3$. These parameters can be
  estimated within a global least squares fit, together with a number
  of parameters of interest, such as post-Newtonian parameters
  $\gamma$ and $\beta$, and the orbits of Mercury and the Earth. The
  simulations have been performed by means of the ORBIT14 orbit
  determination software, developed by the Celestial Mechanics Group
  of the University of Pisa for the analysis of the BepiColombo MORE
  radio science experiment. We claim that the torsion parameters can
  be determined by means of the relativity experiment of BepiColombo
  at the level of some parts in $10^{-4}$, which is a significant
  result in order to constrain gravitational theories which allow
  space-time torsion.
\end{abstract}
\maketitle
%\tableofcontents

\section{Introduction}

The Einstein's General Theory of Relativity (GR) accounts for all the
experimental findings concerning gravitational interaction available
at present. According to GR, the flat Minkowsky spacetime is deformed
by the matter, giving rise to a non-flat and dynamic spacetime, tied
by Riemannian geometry. On the other hand, it is well-known that the
other fundamental interactions can be successfully described, at
microscopic level, within a rigid and flat spacetime. Since GR was
originally formulated as a theory involving mass distribution at
macroscopic level, it is desirable to consider suitable
generalizations of GR that include micro-physical processes, which could
possibly induce macroscopic effects, to be constrained, in turn, by
experiments (cfr., e.g., the discussion in \cite{Hehl76}).

In the following, we will consider a particular generalization of GR
which includes a non-vanishing spacetime torsion. Elementary
particles, which constitute matter, are characterized by their
intrinsic spin. Since spin averages out at macroscopic level, GR
considers that the dynamical behavior of a macroscopic distribution of
mass can be described by the energy-momentum tensor of matter alone,
which is coupled to the metric $g_{\mu\nu}$ of a Riemann
spacetime. However, at microscopic level also the spin angular
momentum plays a role in the dynamics, thus it must be coupled in some
way to spacetime. This fact leads to the formulation of a more general
spacetime, the four-dimensional Riemann-Cartan spacetime,
characterized by an additional, non-Riemannian part of the affine
connection, called the contorsion tensor, which should be coupled to
the spin (see, e.g., \cite{Hehl76,Hehl07}).

Torsion as the anti-symmetric part of an asymmetric affine connection
was firstly introduced one century ago by \'Elie Cartan \cite{Cartan},
who glimpsed the link between spacetime torsion and intrinsic angular
momentum of matter (for an historical perspective see, e.g.,
\cite{Hehl76}). If torsion is admitted, it might affect spinning
particles and thus, indirectly, act on light and test particles
throughout the field equations, which determine the metric. Most of
the torsion theories predict a negligible amount of torsion in the
solar system \cite{Hammond}. Indeed, beyond GR,
theories of gravitation usually assume that torsion couples only to
the intrinsic spin of particles and not to rotational angular momentum
(see, e.g., \cite{Ciuf}).

Recently, Mao et al. \cite{Mao} considered
the issue from another point of view: we can consider a given
experiment in the solar system and model the problem by means of a
non-standard gravitational Lagrangian, which includes a detectable
torsion signal; then, whether the model is in agreement or not with
the data can be experimentally tested. Later on, March et
al. \cite{March} developed a more general framework, within the scope
of the parameterized post-Newtonian (PPN) approximation (see, e.g.,
\cite{Will2018}), in order to test the possible effects due to torsion
around massive bodies in the solar system.

In this perspective, the radio science experiment on-board the
BepiColombo mission to Mercury gives an intriguing opportunity to test
possible modifications of GR in the solar system. BepiColombo is an
ESA/JAXA mission launched in October 2018. The mission aims at a
comprehensive exploration of the planet Mercury, thanks to two
spacecraft, whose orbit insertion around the planet is expected for
the end of 2025 \cite{Benk}. The Mercury Orbiter Radio science
Experiment (MORE) is one of the on-board experiments, devised to
enable a better understanding of the geodesy and geophysics of
Mercury, on one side, and of fundamental physics, on the other (see,
e.g., \cite{Mil01,Mil02,Iess09}). Thanks to full on-board and
on-ground instrumentation capable to perform very precise tracking
from the Earth, MORE will have the chance to determine with very high
accuracy the Mercury-centric orbit of the spacecraft and the
heliocentric orbit of Mercury and the Earth.  Taking advantage from
the fact that Mercury is the best-placed planet to investigate the
gravitational effects of the Sun, MORE will allow an accurate test of
relativistic theories of gravitation (relativity experiment). The test
consists in constraining simultaneously the value of a number of
post-Newtonian (PN) parameters together with some other parameters of
general interest, by means of a global non-linear least squares
fit. This can be achieved, as will be described later, by means of a
dedicated orbit determination software, ORBIT14, developed at the
University of Pisa.

The paper is organized as follows. In Section \ref{sec:2} we briefly
describe the MORE relativity experiment: we present the dynamical
model and we discuss the adopted mathematical methods, together with
some generalities on the orbit determination (OD) software,
ORBIT14. In Section \ref{sec:3} we describe the generalized dynamical
model including the possible effects due to spacetime torsion and we
illustrate how it has been implemented in the software. In Section
\ref{sec:4} we present and discuss the results of the numerical
simulations of the MORE relativity experiment including torsion
parameters. Finally, in Section \ref{sec:5} we present a discussion and our conclusions.

\section{MORE relativity experiment}
\label{sec:2}

It is a long-established fact that a space mission to Mercury, being
the inner planet of the solar system, hence the most sensitive to the
gravitational effects of the Sun, would improve significantly the
determination of the PN parameters, constraining more tightly their
agreement, or disagreement, with GR predictions \cite{Bender}. The BepiColombo
opportunity precisely suits this scope, since it is equipped with
on-board instrumentation capable of very precise tracking from the
Earth \cite{Iess01} in order to perform a comprehensive radio science
experiment (MORE), consisting of a gravimetry, a rotation and a
relativity experiment.

The global experiment aims at determining the following quantities of
general interest:
\begin{itemize}
\item the spherical harmonics coefficients of Mercury gravity field
  (see, e.g., \cite{milgro} - Chap. 13) with a signal-to-noise ratio
  better than 10 up to, at least, degree and order 25 and the Love number
  $k_2$ \cite{Kozai};
\item the parameters defining the model of Mercury's rotation (all the details
  can be found in \cite{Cic12});
\item the ``relativity'' parameters, which are the PN parameters
  $\gamma$, $\beta$, $\alpha_1$, $\alpha_2$ and the Nordtvedt
  parameter $\eta$, which characterize the expansion of the spacetime
  metric in the limit of slow motion and weak field (see, e.g.,
  \cite{Will2018,Will2014}), together with some related parameters, such as
  the oblateness of the Sun $J_{2\odot}$, the solar gravitational mass
  $\mu_{\odot}=GM_{\odot}$ (where $G$ is the gravitational constant
  and $M_{\odot}$ the mass of the Sun), possibly its time derivative
  $\zeta=1/\mu_{\odot}d\mu_\odot/dt$, and the solar angular momentum
  $GS_{\odot}$ which appears in the Lense-Thirring effect on the orbit
  of Mercury (see \cite{addr} for a detailed discussion).
\end{itemize}

To achieve the challenging scientific goals of MORE, it is mandatory
to perform a very precise determination of the orbit of the spacecraft
around Mercury and of the orbit of Mercury and the Earth (replaced, in
fact, with the Earth-Moon barycenter (EMB) for the purposes of our
analysis). This is enabled, in turn, by state-of-the-art
on-board and on-ground instrumentation \cite{Iess01,asmar05}. The
on-board transponder will be able to collect the radio tracking
observables (range, range-rate) up to a goal accuracy (in Ka-band) of
about $\sigma_r =15$ cm at 300 s for one-way range and
$\sigma_{rr}=1.5\times 10^{-4}$ cm/s at 1000 s for one-way range-rate.

\subsection{The heliocentric dynamics of Mercury and the EMB}
\label{sec:2:1}

To perform the OD of the spacecraft, Mercury and the EMB at the
required level of accuracy, their dynamics need to be modeled very
carefully.
%On the one hand, the spacecraft orbit around Mercury is
%expected to have a periodicity of about 2.3 h, while, on the other
%hand, the motion of Mercury and the Earth around the Sun takes place
%over 88 days and 365 days, respectively. Thus, we can handle
%separately the Mercury-centric dynamics of the spacecraft and the
%heliocentric dynamics of the two planets. Moreover,
In practice, although we are dealing with a unique global experiment
and a unique set of measurements, we can conceptually separate the
gravimetry-rotation experiments from the relativity experiment (see,
e.g., \cite{GG}). As a matter of fact, gravimetry and rotation mainly
involve short period phenomena, being preferentially related to the
orbital motion of the spacecraft around Mercury (which has a
periodicity of 2.3 hours), whereas relativistic phenomena take place
over longer time scales, of the order of months or years, thus they
can be analyzed mainly by studying the heliocentric motion of Mercury
and the Earth, taking place over 88 and 365 days,
respectively. Furthermore, the chance to perform the relativity
experiment independently from the others, for the purpose of
simulations, is even more legitimate if we consider the goal
accuracies of the observations: properly comparing $\sigma_r$ and
$\sigma_{rr}$ over the same integration time according to Gaussian
statistics, we find that $\sigma_r/\sigma_{rr}\sim 10^5$ s. Thus, we
can infer that range observations are more accurate when observing
phenomena with periodicity longer than $10^5$ s, like relativistic
phenomena, while the opposite holds for gravimetry and rotation, which
are, in turn, performed mainly by means of range-rate observations.

All the details concerning the Mercury-centric dynamical model of the
spacecraft can be found in \cite{Cic16,GG}. In the following, we will
focus on the heliocentric dynamics of Mercury and the EMB, which represent,
in fact, the core of the MORE relativity experiment.
Adopting a Lagrangian formulation, the equation of motion for Mercury
and EMB are given by the Eulero-Lagrange equations:
\begin{equation}
  \frac{d}{dt}\left ( \frac{\partial L}{\partial\mathbf{v}_i}\right )
  = \frac{\partial L}{\partial\mathbf{r}_i}\,,
  \label{Lag}
\end{equation}
where $\mathbf{r}_i$, $\mathbf{v}_i$ are position and velocity,
respectively, of the $i$-th body ($i=1$ for Mercury, $i=2$ for EMB) in
the Solar System Barycenter (SSB) reference frame and $L$ is the
Lagrangian of the problem. The Lagrangian can be decomposed, in turn,
as
\begin{equation*}
L=L_{New} + L_{GR} + L_{PPN}\,,
\end{equation*}
where $L_{New}$ is the Lagrangian of the Newtonian $N$-body problem,
$L_{GR}$ is the correction due to GR in the PN limit and $L_{PPN}$
includes the contribution due to PN and related parameters. In
particular, the term $L_{PPN}$ describes how each
parameter individually affects the dynamics and it assumes the form:
\begin{equation*}
  L_{PPN}  =  (\gamma -1)L_\gamma + (\beta -1)L_\beta + \eta L_\eta +\alpha_1 L_{\alpha_1} +  \alpha_2 L_{\alpha_2} + J_{2\odot}L_{J_{2\odot}} +\zeta L_\zeta\,.
\end{equation*}
The explicit expression of each term can be found in \cite{GG}. Let us
recall that $\gamma$ and $\beta$ are expected to be equal to unit in
GR, while $\eta$, $\alpha_1$ and $\alpha_2$ are all zero in GR.

From Equation (\ref{Lag}), the total acceleration acting on the $i$-th
body can be derived, taking into account that the main term is the
$N$-body Newtonian acceleration, $\mathbf{a}_i^{New}$, while the other
terms are small perturbations, and it takes the following approximated
expression:
\begin{equation}
  \mu_i\mathbf{a}_i  =  \mu_i\mathbf{a}_i^{New} + \frac{\partial (L-L_{New})}{\partial\mathbf{r}_i}-  \left [\frac{d}{dt}\left (\frac{\partial (L-L_{New})}{\partial\mathbf{v}_i}  \right )  \right ] |_{\mathbf{a}_i=\mathbf{a}_i^{New}} - \ddot{\mathbf{B}}\,,  \label{mai} 
\end{equation}
where $\ddot{\mathbf{B}}$ accounts for the acceleration of the SSB
(see \cite{GG} for details).

\subsection{Mathematical methods}
\label{sec:math}

The parameters of interest for the MORE relativity experiment are
determined simultaneously by means of a global non-linear least squares (LS)
fit.  Following, e.g., \cite{milgro} - Chap. 5, the non-linear LS fit
aims at determining a set of parameters, $\mathbf{u}$, which minimizes
the target function:
\begin{equation*}
Q(\mathbf{u}) = \frac{1}{m}\,\boldsymbol{\xi}^T(\mathbf{u})W\boldsymbol{\xi}(\mathbf{u})\,,
\end{equation*}
where $m$ is the number of observations, $W$ is the matrix containing
the observation weights and $\boldsymbol{\xi} (\mathbf{u}) =
\mathcal{O} - \mathcal{C} (\mathbf{u})$
is the vector of the
residuals, that is the difference between the observations
$\mathcal{O}$ (i.e., the tracking data) and the predictions
$\mathcal{C}(\mathbf{u})$, resulting from the light-time computation
as a function of all the parameters $\mathbf{u}$ (see \cite{Tommei10}
for all the details).

The procedure to compute the set $\mathbf{u}^\star$ of the parameters
which minimizes $Q$ is based on a modified Newton's method called
\emph{differential correction method}. First, we define the design matrix
$B$ and the normal matrix $C$ as
\begin{equation}
B = \frac{\partial \boldsymbol{\xi}}{\partial \mathbf{u}} (\mathbf{u})\,,\,\,\,\,\, C = B^TWB\,.
\end{equation}
 The stationary points of the target function are the solution of the
 normal equation:
\begin{equation*}
C\,\Delta\mathbf{u}^\star = −B^T W \boldsymbol{\xi}\,,
\end{equation*}
where $\Delta\mathbf{u}^\star =\mathbf{u}^\star -\mathbf{u}$. The
method consists in applying iteratively the correction
\begin{equation*}
  \Delta \mathbf{u} = \mathbf{u}_{k+1}-\mathbf{u}_k= -C^{-1}B^TW\boldsymbol{\xi}
\end{equation*}  
until convergence. From a probabilistic point of view, the inverse of
the normal matrix, $\Gamma=C^{-1}$, can be interpreted as the
covariance matrix of the vector $\mathbf{u}^\star$ (see, e.g.,
\cite{milgro} - Chap. 3), carrying information on the attainable
accuracy of the estimated parameters.

\subsection{The ORBIT14 software}

Since 2007, the Celestial Mechanics Group of the University of Pisa
has developed (under an Italian Space Agency agreement) an orbit
determination software, ORBIT14, dedicated to the BepiColombo and Juno
radio science experiments (see, e.g., \cite{Tommei15,Serra16}). All
the code is written in Fortran90.

In the case of BepiColombo, the software includes two separated
stages:
\begin{itemize}
  \item the data simulator: awaiting for real data, it generates the
    simulated observables and the nominal value for the orbital
    elements of the Mercury-centric orbit of the spacecraft and the
    heliocentric orbits of Mercury and the EMB;
\item the differential corrector: it is the core of the code, solving
  for the parameters of interest by means of a global non-linear LS fit, within
  a constrained multi-arc strategy (\cite{Alessi12}).
\end{itemize}
A comprehensive view of the structure of the software can be found in,
e.g., \cite{GG}. The software was successfully used for the analysis
of the real Doppler data of the NASA Juno mission \cite{Deni19}.

\section{Dynamical model with torsion}
\label{sec:3}

Following \cite{March}, we will consider a class of gravitational
theories allowing for non-vanishing torsion based on a Riemann-Cartan
spacetime. This spacetime is a four-dimensional manifold endowed with
a Lorentzian metric $g_{\mu\nu}$ and an affine connection
$\Gamma^\lambda\,_{\mu\nu}$ such that $\nabla_\lambda g_{\mu\nu}= 0$
(mathematical details can be found, e.g., in \cite{Hehl76}). The
connection is uniquely determined by the metric and by the torsion
tensor:
\begin{equation*}
S_{\mu\nu}\,^{\lambda}\equiv \frac{1}{2}(\Gamma^\lambda\,_{\mu\nu}-\Gamma^\lambda\,_{\nu\mu})\,.
\end{equation*}
In this case, the connection departs from the Levi-Civita connection
(which is symmetric in the first two indices and holds in the Riemann
spacetime of GR) by an additional anti-symmetric term, called the
contorsion tensor, $K_{\mu\nu}\,^{\lambda}$, which cancels out in case
of vanishing torsion tensor.

\subsection{Spacetime with torsion in a PPN framework}

In order to develop a model to test torsion within a PPN framework
consistent with the dynamical model described in Section
\ref{sec:2:1}, the metric and the torsion tensor need to be
parameterized in a region of space at a distance $r$ from the Sun such
that the quantity $\epsilon_{\mu_\odot}=\mu_\odot/r\ll 1$, i.e., at large
distances compared with the Schwarzschild radius. Since the PN
framework we are adopting includes terms up to second order in the
small quantity $\epsilon_{\mu_\odot}$ (see, e.g., \cite{Mao,March}), the metric
and the torsion tensor must be expanded up to the same degree of
accuracy.

In general, the torsion tensor has 24 independent components, each
being a function of time and position, but, adopting symmetry
arguments and a perturbative approach up to
$O^2(\epsilon_{\mu_\odot})$, March et al. \cite{March} showed that ultimately there
are only 2 independent components of the torsion tensor, which can be
parameterized by means of 3 independent parameters, the torsion
parameters $t_1$, $t_2$, $t_3$.

In a Riemann-Cartan spacetime two different classes of curves can be
considered, autoparallels and geodesics curves, which reduce both to the
geodesics of the Riemann spacetime when torsion vanishes
\cite{Hehl76}. In particular, autoparallels are curves along which the
velocity vector is transported parallel to itself by the connection
$\Gamma^\lambda\,_{\mu\nu}$, while along geodesics the velocity vector
is transported parallel to itself by the Levi-Civita connection. In GR
the two types of trajectories coincide while, in general, they may
differ in the presence of torsion.  Since geodesics curves in this
generalized framework turn out to be the same as in the standard PPN
framework (see \cite{March} for details), new predictions related to
torsion may arise only when considering autoparallel trajectories,
which explicitly depend on torsion parameters.

After some mathematical manipulations and simplifications and
accounting only for terms up to the second order in the small quantity
$\epsilon_{\mu_\odot}$, March et al. \cite{March} achieved the final equation of
motion for a test body moving along an autoparallel trajectory,
written in rectangular coordinates:
\begin{equation}
 \ddot{x}^\alpha  =  -m\frac{x^\alpha}{r^3} + 2[\beta+t_1(1+\gamma)-t_3]\, m^2\frac{x^\alpha}{r^4} +  (t_1+t_2 -2 )\,m\frac{\dot{x}^\alpha}{\dot{r}}{r^2} -   (2\gamma + t_2)\,m\frac{x^\alpha}{r^3}v^2 + 3\gamma\, m\frac{x^\alpha}{r^3}\dot{r}^2\,, \label{x2alpha}
\end{equation}
where $\alpha\in \{1,2,3\}$, $v^2=\sum_{\alpha=1}^3(\dot{x}^\alpha)^2$
and $m$ is the mass of the central massive body ($G=c=1$ has been
assumed)\footnote{We rearranged Equation (6.6) in
  \cite{March}, making explicit the dependence of the coefficients by
  $(\gamma,\,\beta,\,t_1,\,t_2,\,t_3)$.}. We also recall that,
differently from our approach, in \cite{March} all the PN parameters
other than $\gamma$ and $\beta$ are assumed to be zero, hence
they are not included in the development and parameterization of the
metric and the connection.

\subsection{Implementation of torsion in ORBIT14}
\label{impl}

To properly account for the possible dynamical effects due to
spacetime torsion, we need to add the torsion contribution to the
global acceleration acting on the $i$-th body, described by
Equation~({\ref{mai}}). In principle, we could directly use
Equation~(\ref{x2alpha}), but, for the purposes of the MORE relativity
experiment, we need to take into account the possible dynamical
effects due to all the PN and related parameters we are interested in,
since they all need to be determined simultaneously in the LS
fit. Nevertheless, assuming that the only contribution to the
Lagrangian $L_{PPN}$ is due to $\gamma$ and $\beta$,
Equations~(\ref{mai}) and (\ref{x2alpha}) must coincide apart from the
terms due to torsion. This, in fact, is the case; thus, after
some manipulation, we can isolate the contribution due to torsion on
the motion of the $i$-th body, which reads \footnote{We omitted the
  multiplicative factor $1/c^2$, while we restored the $G$ factor.}:
\begin{equation}
  \mathbf{a}_i^{tor} =  2[t_1 (1+\gamma) - t_3] \,\mu_\odot^2\, \frac{\mathbf{r}_{i,\odot}}{r_{i,\odot}^4} +   (t_1+t_2)\, \mu_\odot \, \frac{\mathbf{r}_{i,\odot}\cdot\mathbf{v}_{i,\odot}}{r_{i,\odot}^3}\,\mathbf{v}_{i,\odot} 
   -  t_2\,\mu_\odot\, \frac{\mathbf{v}_{i,\odot}\cdot\mathbf{v}_{i,\odot}}{r_{i,\odot}^3}\,\mathbf{r}_{i,\odot}\,, \label{ator}
\end{equation}
where $\mathbf{r}_{i,\odot}\equiv \mathbf{r}_i-\mathbf{r}_\odot$ is
the position of the body $i$ with respect to the SSB and
$r_{i,\odot}=|\mathbf{r}_{i,\odot}|$ is its module (the same notation
holding for the velocity vector).

Two issues stand out by looking at Equation~(\ref{ator}). The first
one concerns the fact that the contribution to the acceleration due to
each of the three torsion parameters depends on the gravitational mass
of the Sun, $\mu_\odot$, which is one of the parameters to be
determined as well; in the case of $t_1$, also a coupling with
$\gamma$ occurs. Thus, we need to properly take into account this
dependency in the computation of the design and normal matrices, which
must contain the partial derivatives of the new acceleration term with
respect to $\mu_{\odot}$ and $\gamma$. As a consequence, we can expect
that a considerable correlation between $\mu_{\odot}$, $\gamma$ and
the torsion parameters could show up by solving simultaneously for all
the parameters.

The second issue concerns specifically $\beta$ and $t_3$. Looking at
Equation~(\ref{x2alpha}), it turns out that the dynamical effect due
to these two parameters shows the same proportionality to the term
$\mathbf{r}_{i,\odot}/r_{i,\odot}^4$. Thus, we can define an overall
acceleration term due to the combined effect of both parameters, given
by:
\begin{equation}
  \mathbf{a}_i^{\beta -t_3} = 2\,(\beta -t_3)\, \mu_\odot^2 \frac{\mathbf{r}_{i,\odot}}{r_{i,\odot}^4}\,.
  \label{beta-}
\end{equation}
As a consequence, in principle we cannot solve simultaneously for the
two parameters within the LS fit, since the signatures due to each of
the two effects on the observations cannot be distinguished in any
way. From a computational point of view, this means that the design
matrix would have two linearly dependent columns, so that the normal
matrix would be degenerate, and therefore not invertible. This
circumstance is usually known as rank deficiency and a general
description of the issue in the case of an OD problem can be found in
\cite{milgro} - Chap. 6. Similar issues have been already dealt with in
the past in the framework of the MORE relativity experiment (a
detailed discussion on the approximate rank deficiency between $\beta$
and the oblateness of the Sun, $J_{2\odot}$, can be found in
\cite{Mil02}, while the general issue in the case of the MORE
relativity experiment has been widely discussed by the authors in
\cite{addr,Serra16}).

Rank deficiencies in an OD problem can be cured in two ways: solving
for fewer parameters (this technique is known as descoping) or using
more observations. The second way can be pursued by adding independent
observations (e.g., using data from other experiments \cite{DeM20})
or, similarly, by adding a number of constraints equal to the order of
the rank deficiency, which represent the available apriori knowledge
of the problem (see, e.g., \cite{addr}).  The first approach
(descoping) has been applied, in fact, in the framework of the
relativity experiment until now: assuming a vanishing torsion, as in
GR, is equivalent to assuming that $t_3=0$; thus, looking at
Equation~(\ref{beta-}), the actual effect on the dynamics is all due
to the parameter $\beta$. This will be the strategy applied in the
scenarios of the reference simulation and of simulation (a) described
in Section \ref{sim_res}.  On the other hand, if we consider that a
non-vanishing torsion may exist, and, in particular, that $t_3$ could
be not null, the descoping approach cannot be adopted anymore. In this
case we cannot separate the effects of $\beta$ and $t_3$ and we can
only estimate the linear combination $(\beta- t_3)$. In we want to
determine the two parameters, we need to add some further information
on, at least, one of the two parameters. This can be done by adding an
apriori constraint on the parameter $\beta$, given by the present
knowledge on its value. If the apriori constraint is sufficiently
tight, then the normal matrix can be inverted again and the problem
can be solved. This approach will be adopted in the case of simulation
(b) in Section \ref{sim_res}.

\section{Numerical analysis}
\label{sec:4}

In this Section, we describe the results of the numerical simulations
of the MORE relativity experiment, including the dynamical effects due
to torsion. In particular, we aim at checking whether the torsion
parameters can be estimated by means of the experiment and, if so, at
which level of accuracy.

\subsection{Simulation scenario and assumptions}
\label{sec4_1}

First of all, we define a reference scenario. In this case, the LS fit
aims at determining the relativity parameters listed in Section
\ref{sec:2}, without accounting for possible effects due to
torsion in the solution (see, e.g., \cite{addr}). We recall that the relativity parameters are: the PN
parameters $\beta$, $\gamma$, $\alpha_1$, $\alpha_2$, the Nordtvedt
parameter $\eta$, the oblateness of the Sun $J_{2\odot}$, the
gravitational mass of the Sun $\mu_\odot$, its time derivative $\zeta$
and the angular momentum of the Sun $GS_\odot$. 

The starting epoch of the simulation is set to 14 March 2026 and the
nominal duration is set to 1 year. In Section \ref{benef} we will
consider also possible benefits due to an extension of the mission
duration up to an additional year. We assume that the radio tracking
observables are affected only by random effects with a standard
deviation of $\sigma_r =15$ cm at 300 s and $\sigma_{rr}= 1.5\times
10^{-4}$ cm/s at 1000 s, respectively, for Ka-band observations. The
software is capable of including also a possible systematic component
to the range error model and to calibrate for it, but, since we are
interested in a covariance analysis, we do not account for this
detrimental effect here (this issue has been partially discussed in
\cite{metro16}).

For each relativity parameter we are interested in, we set an apriori constraint given by
the present knowledge of the parameter, as listed in Table
\ref{tab_ref}. We point out that the adopted apriori on $\gamma$ is a
conservative estimate derived by a full set of simulations, carried
out by the authors in \cite{serra19}, of the Superior Conjunction
Experiment (SCE), planned during the cruise phase of
BepiColombo \footnote{The first solar superior conjunction of
  BepiColombo is expected on March 2021.}. Concerning the torsion
parameters, they have not been experimentally
estimated so far. In \cite{March}, referring to the present knowledge
of $\beta$ and $\gamma$, the authors derived the following constraints
on the values of $t_2$ and $t_3$:
\begin{equation*}
|t_2| < 0.0128\,,\;\;\;\;\; |t_3| < 0.0286\,.
\end{equation*}

Moreover, we make use of an important assumption: we link the PN
parameters by the Nordtvedt equation \cite{nordt}
\begin{equation*}
  \eta = 4(\beta -1) -(\gamma -1) -\alpha_1 -\frac{2}{3}\alpha_2\,,
\end{equation*}
which means that we are considering only metric theories of
gravitation. The addition of the Nordtvedt equation in our model is
motivated by the fact that $\beta$ and $J_{2\odot}$ are expected to
show an almost 100\% correlation, since their dynamical effect on the
orbit of Mercury is comparable (see, e.g., \cite{Mil02} for an
extensive discussion of this symmetry in the case of the MORE
relativity experiment). Thus, the addition of the constraint removes
the degeneracy between $\beta$ and $J_{2\odot}$, but this result is,
in turn, obtained at the cost of forcing an almost 100\% correlation
between $\beta$ and $\eta$ \footnote{This follows from the fact that,
  since $\gamma$ is highly constrained by the SCE apriori and
  $\alpha_1$ and $\alpha_2$ can be neglected, the Nordtvedt equation
  forces a linear dependency of $\eta$ from $\beta$.}.

\begin{table}
  \begin{ruledtabular}
          \caption{Present knowledge of the relativity parameters.}\label{tab_ref}
          \begin{tabular}{cccc}
            Parameter & Accuracy & Parameter & Accuracy \\
            \hline
    $\beta$ & $3.0\times 10^{-5}$ \cite{fienga} &  $J_{2\odot}$ &  $1.2\times 10^{-8}$ \cite{fienga} \\
    $\gamma$ &  $1.0\times 10^{-5}$ \cite{serra19} & $\mu_{\odot}$ &  $3.0\times 10^{14}$ \footnote{Value from latest JPL ephemerides publicly available at: http://ssd.jpl.nasa.gov/?constants. Accessed 23 June 2020.} \\
   $\eta$ & $4.4\times 10^{-4}$ \cite{williams}  &  $\zeta$ &  $1.0\times 10^{-14}$ \cite{pit} \\
    $\alpha_1$ & $6.0\times 10^{-6}$ \cite{iorio14} & $GS_{\odot}$  & $1.3\times 10^{40}$ \cite{park} \\
    $\alpha_2$ & $3.5\times 10^{-5}$ \cite{iorio14} &   & \\
    \hline        
  \end{tabular}
  \end{ruledtabular}
\end{table}

\subsection{Simulation results}
\label{sim_res}

In the following, we denote as ``reference simulation'' the case in
which the solution is computed in the standard reference scenario
described in Section \ref{sec4_1}, which consists in estimating the
state vectors of Mercury and the EMB at the central epoch of the
mission ($6+6$ parameters) and the relativity parameters, for a total
of 21 parameters. Then, we define as
simulation (a) the case of the reference scenario with the addition of
the torsion parameters $t_1$ and $t_2$ in the solve-for list, for a
total of 23 estimated parameters, and we estimate $(\beta-t_3)$ in
place of $\beta$. Finally, we label as simulation (b) the case of a
simulation in which all the three torsion parameters, $t_1$, $t_2$,
$t_3$, are simultaneously determined by means of the LS fit together
with the relativity parameters and the state vectors, for a total of
24 parameters. In addition, in each scenario the state vector of the
spacecraft (position and velocity, 6 parameters) is estimated at the
central epoch of each of the 365 observed arcs, for a total of 2190
parameters. These parameters are handled as local
parameters (i.e., they belong only to a single arc: see, e.g.,
\cite{Alessi12} for details) and they result uncorrelated with the
relativity parameters, which are global parameters. For the sake of
clarity, the global parameters estimated in each of the
three scenarios are summarized in Table \ref{lista}.
\begin{table}
  \begin{ruledtabular}
    \caption{Summary of the parameters solved in the LS fit in each of the three scenarios described in the text.}\label{lista}
  \begin{tabular}{ll}
   Scenario & Solved parameters  \\
  \hline
  Reference & state vectors of Mercury and EMB; relativity parameters \\
  Simulation (a) & state vectors of Mercury and EMB; relativity parameters; $t_1$, $t_2$ \\
  Simulation (b) & state vectors of Mercury and EMB; relativity parameters; $t_1$, $t_2$, $t_3$ \\
  \end{tabular}
  \end{ruledtabular}
\end{table}

%At this stage, we need to point out that each of the three scenarios
%employs as input the same simulated data, which have been generated
%assuming that the torsion parameters, as the N parameters, are null (as expected in
%GR). Nevertheless, we recall that a different assumption at simulation
%stage (i.e., non-null value of torsion parameters) would not change
%the results of the differential correction stage in terms of formal accuracies determined by the LS
%fit. Indeed, we are interested, for the moment, in determining at
%which level a parameter can be constrained by the experiment, not in
%determining the absolute value of the parameter, which, instead, will
%become a crucial issue when we will deal with real
%data. \textcolor{red}{Non so se il discorso sia chiaro, mi sembrava
%  che avesse senso specificare questa cosa.}

The results in terms of formal accuracies (1-$\sigma$) determined in
the three different scenarios are compared in Table \ref{tab_res2} for
what concerns the state vectors (position and velocity in cartesian
coordinates) of Mercury (labeled with the index $M$) and the EMB
(labeled with the index $E$), while in Table \ref{tab_res1} the
corresponding results for the relativity parameters are shown.

\begin{table}
  \begin{ruledtabular}
    \caption{Formal accuracies expected for the state vectors
      (position and velocity in cartesian coordinates) of Mercury
      (index $M$) and the EMB (index $E$) in the case of: reference
      simulation, simulation (a) and simulation (b), respectively. In
      the last column, we introduce a reference number, $N$, which
      labels each parameter, that will be adopted in Section \ref{sec_corr}. The
      results are expressed in cm and cm/s.}\label{tab_res2}
  \begin{tabular}{ccccc}
   Parameter & Reference  & Simulation (a) & Simulation (b) & $N$ \\
  \hline
  $x_M$ & $4.5\times 10^{2}$ & $4.5\times 10^{2}$ & $4.5\times 10^{2}$ & 1 \\
   $y_M$ & $2.7\times 10^{2}$  & $2.7\times 10^{2}$ & $2.7\times 10^{2}$ & 2 \\
  $z_M$ & $1.6\times 10^{3}$ & $1.6\times 10^{3}$  & $1.6\times 10^{3}$ & 3 \\
  $x_E$ & $5.9\times 10^{1}$  & $5.9\times 10^{1}$ & $5.9\times 10^{1}$ & 4 \\
  $y_E$ & $1.1\times 10^{3}$  & $1.1\times 10^{3}$ & $1.1\times 10^{3}$ & 5 \\
  $z_E$ & $4.3\times 10^{3}$ & $4.3\times 10^{3}$ & $4.3\times 10^{3}$ & 6 \\
  $v_{x,M}$ & $2.8\times 10^{-4}$ & $2.8\times 10^{-4}$ & $2.8\times 10^{-4}$ & 7 \\
  $v_{y,M}$ & $2.6\times 10^{-4}$ & $2.6\times 10^{-4}$ & $2.6\times 10^{-4}$ & 8 \\
  $v_{z,M}$ & $1.0\times 10^{-3}$ & $1.0\times 10^{-3}$ & $1.0\times 10^{-3}$ & 9 \\
  $v_{x,E}$ & $2.2\times 10^{-4}$ & $2.2\times 10^{-4}$ & $2.2\times 10^{-4}$ & 10 \\
  $v_{y,E}$ & $8.1\times 10^{-6}$ & $8.1\times 10^{-6}$  & $8.4\times 10^{-6}$ & 11 \\
  $v_{z,E}$ & $6.2\times 10^{-4}$ & $6.2\times 10^{-4}$ & $6.3\times 10^{-4}$ & 12 \\
  \end{tabular}
  \end{ruledtabular}
\end{table}

\begin{table}
  \begin{ruledtabular}
    \caption{Formal accuracies expected for the relativity parameters
      in the case of: reference simulation, simulation (a) and
      simulation (b), respectively. In the last column, we introduce a
      reference number, $N$, which labels each parameter, that will be
      adopted in Section \ref{sec_corr}. Note that $\sigma (\mu_\odot
      )$ is expressed in cm$^3$/s$^2$, $\sigma (\zeta)$ in y$^{-1}$
      and $\sigma (GS_\odot )$ in cm$^3$/s$^2$.}\label{tab_res1}
  \begin{tabular}{ccccc}
   Parameter & Reference  & Simulation (a) & Simulation (b) & $N$ \\
  \hline
  $\beta$ & $1.5\times 10^{-5}$ & $2.4\times 10^{-5}$ & $2.6\times 10^{-5}$ & 13 \\
   $\gamma$ & $7.6\times 10^{-7}$  & $7.7\times 10^{-7}$ & $7.9\times 10^{-7}$ & 14 \\
  $\eta$ & $6.1\times 10^{-5}$ & $9.8\times 10^{-5}$  & $1.0\times 10^{-4}$ & 15 \\
  $\alpha_1$ & $5.9\times 10^{-7}$  & $6.2\times 10^{-7}$ & $6.2\times 10^{-7}$ & 16 \\
  $\alpha_2$ & $9.8\times 10^{-8}$  & $1.1\times 10^{-7}$ & $1.3\times 10^{-7}$ & 17 \\
    $\mu_{\odot}$ & $6.4\times 10^{13}$ & $1.9\times 10^{14}$ & $2.4\times 10^{14}$ & 18 \\
  $J_{2\odot}$ & $1.9\times 10^{-9}$ & $2.3\times 10^{-9}$ & $2.4\times 10^{-9}$ & 19 \\
  $\zeta$ & $9.1\times 10^{-15}$ & $9.5\times 10^{-15}$ & $1.0\times 10^{-14}$ & 20 \\
  $GS_{\odot}$ & $1.2\times 10^{40}$ & $1.3\times 10^{40}$ & $1.3\times 10^{40}$ & 21 \\
  \hline
  $t_1$ & -- & $1.7\times 10^{-5}$ & $1.9\times 10^{-4}$ & 22 \\
  $t_2$ & -- & $1.3\times 10^{-5}$  & $1.4\times 10^{-4}$ & 23 \\
  $t_3$ & -- & -- & $3.6\times 10^{-4}$ & 24 \\
  \end{tabular}
  \end{ruledtabular}
\end{table}

Concerning the determination of the state vectors at the central
epoch, shown in Table \ref{tab_res2}, we can immediately observe that,
as expected, the results are the same in all scenarios, suggesting
that the there is no correlation between the torsion parameters and
any of the components of the orbital conditions of Mercury and the
EMB, i.e., solving for the torsion parameters does not weaken the determination of the state vectors. The two orbits can be determined with the MORE experiment at the
level of some meters, for what concerns the position, and of some m/s,
concerning the velocity, with a general trend of a better estimate in
the $x-y$ plane, while the $z$-direction seems weaker. We point out
that no constraint has been imposed on the two orbits (contrarily to
what the authors have done in \cite{addr}): the issue of possible
symmetries affecting the geometry of our experiment is very critical
and a dedicated paper will be devoted by the authors to this
specific topic in the near future.

Moving the attention to Table \ref{tab_res1}, first of all some
observations on the results achievable in the case of the reference
scenario are in order. Comparing the estimated accuracy with the
present knowledge, shown in Table \ref{tab_ref}, we can expect a
significant improvement in the knowledge of $\gamma$, $\alpha_1$,
$\alpha_2$ and $J_{2\odot}$ thanks to the MORE experiment. A minor
improvement could be achieved in the estimate of the gravitational
mass of the Sun, $\mu_\odot$, while no significant improvement is
expected, at the moment, for the knowledge of $\zeta$ (see \cite{GG}
for a discussion about this parameter) and $GS_\odot$ (see \cite{addr}
for a further discussion).

Concerning the torsion parameters, which are the focus of this work,
we can observe that in the case of simulation (a) we achieve the
relevant result that the torsion parameters $t_1$ and $t_2$ can be
estimated by means of the MORE relativity experiment at the level of
$10^{-5}$, which could place a significant constraint on the validity
or not of torsion theories. In the case of simulation (b),
i.e. estimating also the parameter $t_3$, an order of magnitude is
lost in the determination of the torsion parameters. This is mainly a
consequence of the strong correlation between $\beta$ and $t_3$: since
the three torsion parameters are expected to be correlated one with
the other (see Equation~(\ref{ator})), the worsening affects the
estimate of all the three parameters. Anyway, they can be estimated at
the level of some parts in $10^{-4}$ and this is still a significant
result.

The addition of the torsion parameters in the solve-for list worsens
the determination of most of the relativity parameters, slightly more
in the case of simulation (b) with respect to simulation (a). The
worsening is more pronounced in the case of the parameter
$\mu_{\odot}$, whose accuracy deteriorates by a factor 3 in simulation
(a) and by a factor 4 in simulation (b). This result is not surprising
since the dynamical effect due to torsion depends on $\mu_\odot$, as
pointed out in Section \ref{impl}. In particular, in the case of
simulation (b), the accuracy of the parameter $\mu_\odot$ is
comparable to the present knowledge and this fact suggests that, in
the absence of an apriori on $\mu_\odot$, it would be difficult to
solve for this parameter simultaneously with the torsion
parameters. In the case of the parameters $\beta$ and $\eta$, the
worsening is limited by less than a factor 2. This fact may seem
contradictory in the case of simulation (b). Indeed, in Section
\ref{impl} we have seen that the dynamical effect due to $\beta$ and
$t_3$ is the same and, as a consequence, we could expect that the
addition of $t_3$ in the solve-for list should cause a significant
worsening in the determination of $\beta$. This is, in fact, the case:
if we do not constrain the value of $\beta$ in some way and we
simultaneously estimate $t_3$, we would find that the normal matrix is
not invertible. The problem is solved by adding the present knowledge
on $\beta$ as an apriori constraint on the parameter: we are not able
to improve the present knowledge of $\beta$ (by looking at Table
\ref{tab_res1} - simulation (b), the obtained formal accuracy is equal
to the present knowledge), but at least we are able to separate its
effect from that of $t_3$.

A special discussion deserves the issue of the determination itself of
$\beta$ and $\eta$. We recall that the two parameters are highly
correlated since they are linked through the Nordtvedt relation. In
the past, we observed that the determination of $\beta$ and $\eta$
strongly depends on the level at which the experiment is capable to
determine the orbits of Mercury and the EMB. In particular, referring
to the results shown in \cite{addr}, we observed that the
determination of the orbits of Mercury and the EMB with an accuracy
below the meter and m/s for the position and velocity, respectively,
leads to a great benefit in the estimate of $\beta$ and $\eta$
(cfr. Tables 2 and 3 in \cite{addr}). The problem is that an accuracy
at the level of 10 cm or less in the determination of the two orbits
seems unrealistic, due to a number of detrimental effects, in
particular due to the uncertainty in the knowledge of the masses and
state vectors of the bodies in the solar system, which influences in
turn the position and velocity of Mercury and the EMB (mainly in the
case of Jupiter, see \cite{DM16} for an extensive discussion). Indeed,
we point out that, for the purpose of the experiment, the mass and
position of the bodies included in the $N$-body Newtonian acceleration
(see Equation~(\ref{mai})) are taken from the JPL ephemerides, thus
they are assumed as perfectly known. As a consequence, we cannot
expect to significantly improve our knowledge on $\beta$ and $\eta$
with the relativity experiment set up in the way described
here. Different choices can be made, considering a wider set of
independent observations due to different space missions to be fitted
simultaneously, as proposed in \cite{DeM20}.

\subsection{Analysis of the correlations}
\label{sec_corr}

To carry out a complete description of the results, the correlations
showing up between the estimated parameters need to be analyzed. We
recall that the correlation of the parameter $A$ with the parameter $B$,
$\rho(A,B)$, where $A$ and $B$ are, respectively, the $i$-th and the
$j$-th parameters in the solve-for list, is proportional to the
out-of-diagonal $\Gamma_{ij}$ term of the covariance matrix $\Gamma$
(see Section \ref{sec:math}) and it is normalized to 1 by means of the
formal accuracy of the two parameters, $\sigma (A)$ and $\sigma (B)$.

\begin{figure}
  \includegraphics[width=0.5\textwidth]{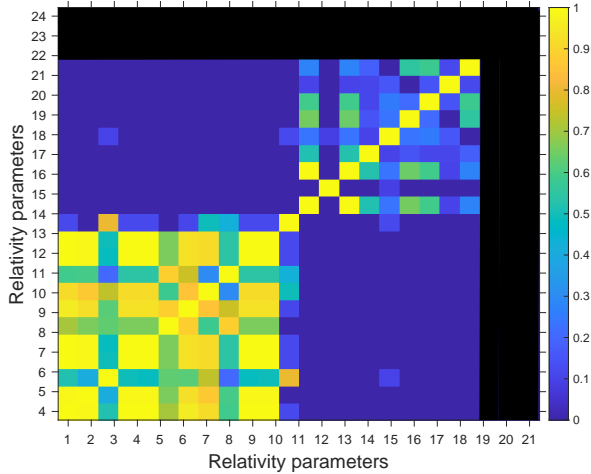}
  \caption{Correlations between the estimated parameters in the case of the reference simulation. Parameters are labeled as in Tables \ref{tab_res2} and \ref{tab_res1}.} \label{fig1}
\end{figure}

\begin{figure}
  \includegraphics[width=0.5\textwidth]{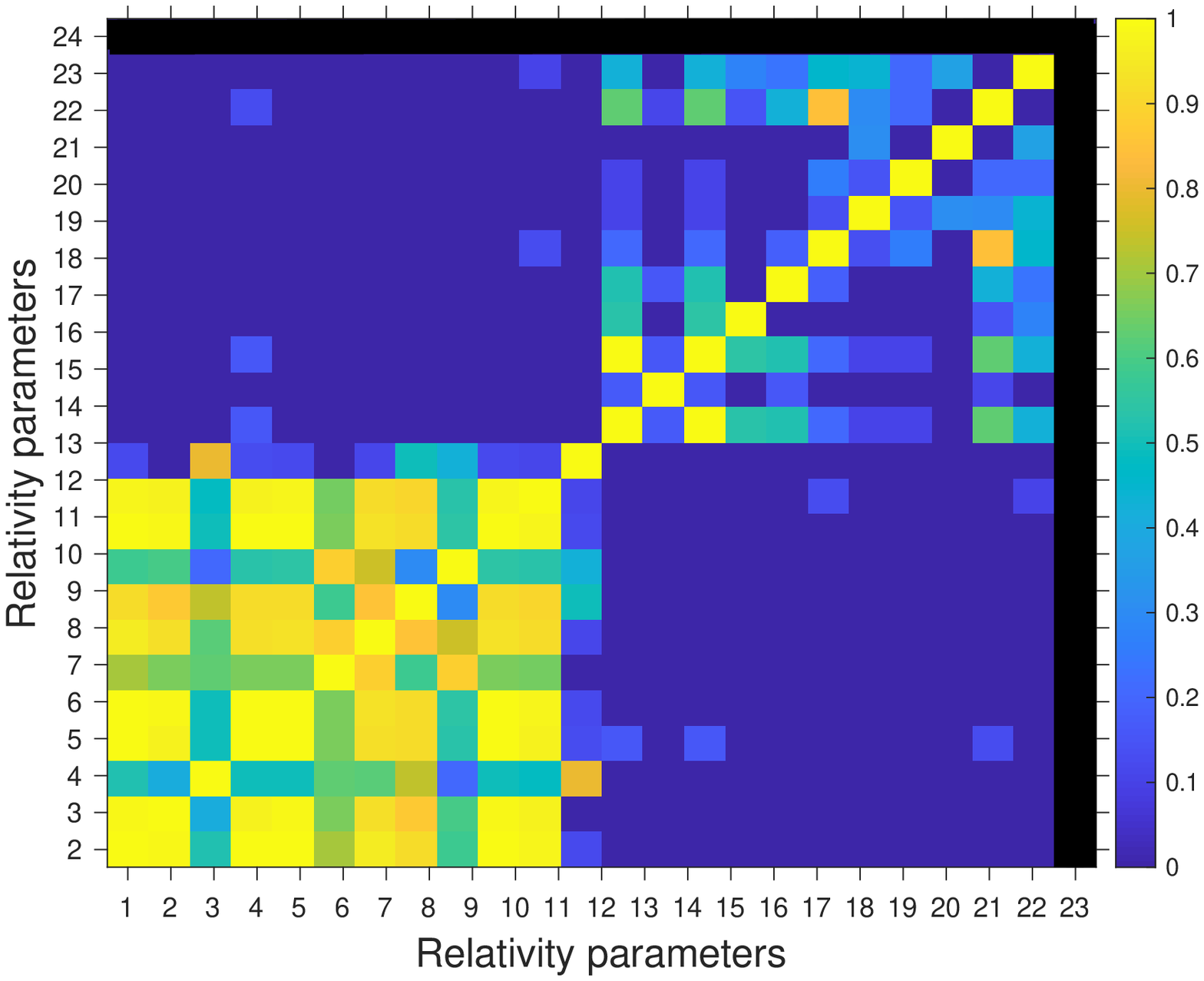}
  \caption{Correlations between the estimated parameters in the case of simulation (a). Parameters are labeled as in Tables \ref{tab_res2} and \ref{tab_res1}.} \label{fig2}
\end{figure}

\begin{figure}
  \includegraphics[width=0.5\textwidth]{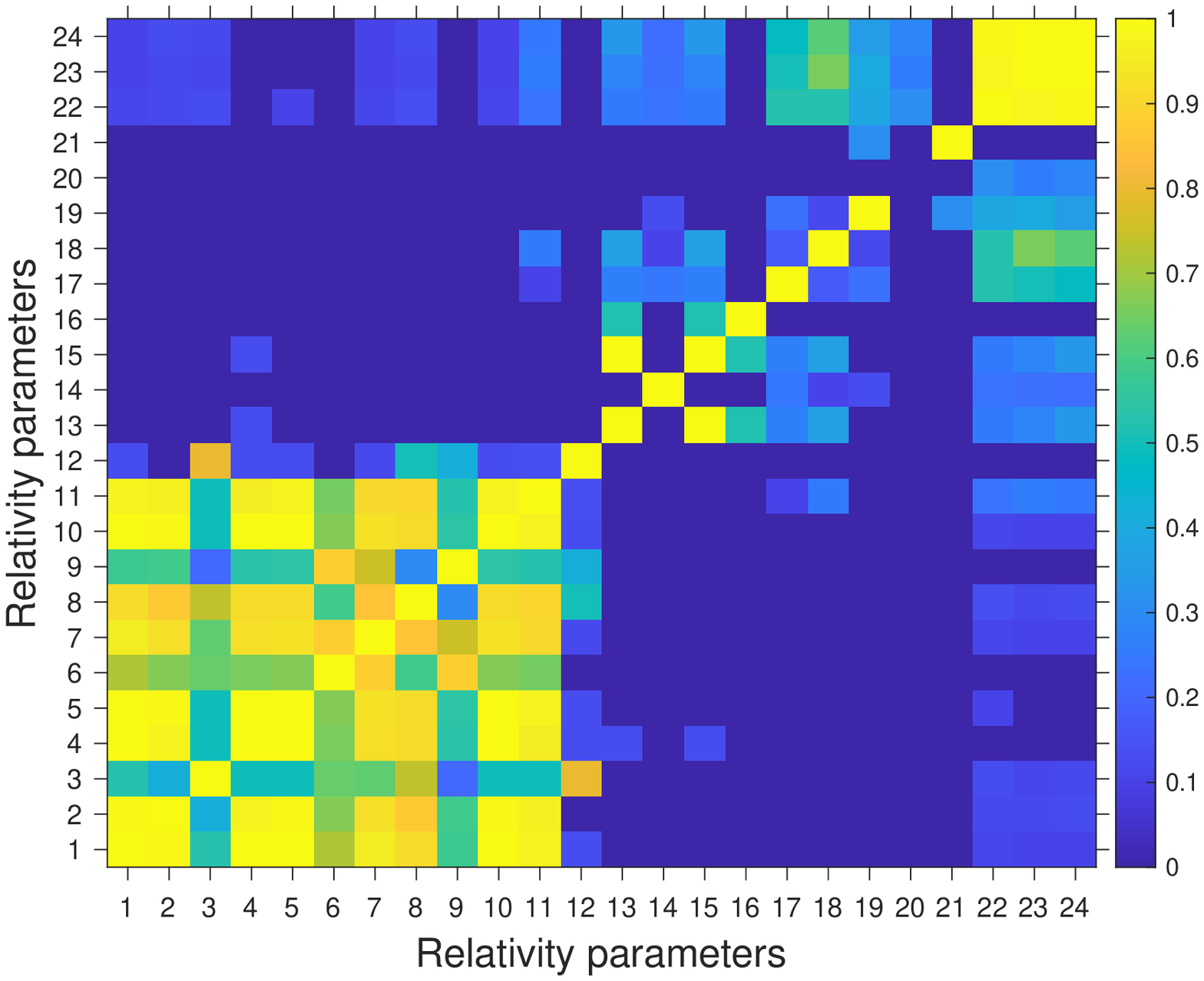}
  \caption{Correlations between the estimated parameters in the case of simulation (b). Parameters are labeled as in Tables \ref{tab_res2} and \ref{tab_res1}.} \label{fig3}
\end{figure}

A global view of the behavior of correlations between the parameters
is given in Figures \ref{fig1}, \ref{fig2} and \ref{fig3} for the
cases of the reference simulation, simulation (a) and simulation (b),
respectively. The estimated parameters are labeled with a number
$N=1,...,24$ as in Tables \ref{tab_res2} and \ref{tab_res1}: indices
from 1 to 12 refer to the state vectors of Mercury and the EMB
(position and velocity components), from 13 to 21 to the relativity
parameters, in the order indicated in Table \ref{tab_res1}, and,
finally, from 22 to 24 to the three torsion parameters. The color-bar
on the right of each figure spans from the value 0 (dark blue) in the
case that no correlation is found to the value 1 (yellow) in the case
that an exact correlation shows up.

In the case of the reference simulation (Figure \ref{fig1}) a
two-block structure can be clearly recognized, which can be found also
in the cases of simulation (a) and (b). This structure reveals the
fact that the subset of parameters 1-12, i.e., the components of the
state vectors, and the subset of the parameters 13-21, i.e., the
relativity parameters, are almost not correlated one with the others,
while one parameter can be highly correlated only with the other
parameters of the same subset. In particular, we can observe that the
correlation between $\beta$ and $\eta$ is $\rho(\beta,\eta)=0.99$ in
all the three cases, as expected. Moreover, $GS_\odot$ shows a
non-null correlation with both $J_{2\odot}$ and $\mu_{\odot}$, equal
to $\rho(J_{2\odot},GS_\odot)=0.58$ and
$\rho(\mu_\odot,GS_\odot)=0.55$. The issue of the correlation showing
up between $J_{2\odot}$ and $GS_\odot$ is described in
\cite{addr}. Finally, $J_{2\odot}$ shows a residual correlation with
$\beta$, and consequently with $\eta$, equal to
$\rho(\beta,J_{2\odot})=0.59$ and $\rho(\eta,J_{2\odot})=0.59$ and
also $\mu_\odot$ turns out to be correlated with these two parameters,
at the level of $\rho(\beta,\mu_\odot)=0.65$ and
$\rho(\eta,\mu_\odot)=0.65$.

In the case of simulation (a), shown in Figure \ref{fig2}, the
behavior of correlations is similar to the reference case. Now that
the parameters $t_1$ and $t_2$ are included in the solve-for list, it
shows up a non-null correlation between $t_1$ and $\mu_{\odot}$, equal
to $\rho(\mu_{\odot},t_1)=0.85$, as it was expected by looking at
Equation~(\ref{ator}). In the case of $t_2$, instead, the correlation
with $\mu_\odot$ is only $\rho(\mu_\odot,t_2)=0.46$. Conversely,
almost no correlation shows up between the torsion parameters and
$\gamma$: this can be ascribed to the presence of a tight apriori on
$\gamma$, as expected from the SCE during the cruise phase. Moreover,
$t_1$ is also partially correlated with $\beta$ and $\eta$, at the
level of $\rho(\beta,t_1)=0.63$ and $\rho(\eta,t_1)=0.63$.

Finally, in the case of simulation (b), shown in Figure \ref{fig3},
the full correlation between the three torsion parameters shows up. If
we look at Equation~(\ref{ator}), we can see that, as long as we assume
$t_3=0$ (simulation (a)), the effects due to $t_1$ and $t_2$ are, at
least partially, independent, thus a low correlation between $t_1$ and
$t_2$ is expected; indeed, we find that their correlation is below
0.1. When also $t_3$ comes into play, since its effect is proportional
to one of the terms due to $t_1$ (see Equation~(\ref{ator})), we find
that $\rho_{(b)}(t_1,t_3)=0.99$ and, as a consequence,
$\rho_{(b)}(t_3,t_2)=0.99$ and $\rho_{(b)}(t_1,t_2)=0.98$. The
correlation between the torsion parameters and the gravitational mass of the Sun, $\mu_\odot$, persists also
in this case, at the level of $\rho(\mu_\odot,t_1)=0.53$,
$\rho(\mu_\odot,t_2)=0.66$ and $\rho(\mu_\odot,t_3)=0.62$,
respectively.

\subsection{Possible benefits from an extended mission}
\label{benef}

The nominal duration of the BepiColombo orbital phase is fixed to one
year, with the scientific operations in orbit starting on March
2026. Anyway, the possibility of an extension up to one further year is
envisaged. Thus, to check the possible benefits due to a 2-year
duration of the experiment, we re-ran the simulations in the three
scenarios described in Section \ref{sim_res}. The results on the
achievable 1-$\sigma$ accuracies for the relativity experiment are shown in
Table \ref{tab_res3}, where for each case we show also the improvement
factor, $R_{\textrm{1-to-2}}$, defined as the ratio of the 1-year accuracy over the 2-year accuracy.
\begin{table*}
  \begin{ruledtabular}
    \caption{Formal accuracies expected for the relativity parameters
      considering a 2-year long orbital experiment, in the case of:
      reference simulation, simulation (a) and simulation (b),
      respectively. For each case we show the 1-$\sigma$ accuracy
      (columns 2, 4, 6) and the improvement factor,
      $R_{\textrm{1-to-2}}$, defined as the ratio of the 1-year over
      the 2-year accuracy (columns 3, 5, 7). Note that $\sigma
      (\mu_\odot )$ is expressed in cm$^3$/s$^2$, $\sigma (\zeta)$ in
      y$^{-1}$ and $\sigma (GS_\odot )$ in
      cm$^3$/s$^2$.}\label{tab_res3}
  \begin{tabular}{ccccccc}
   Parameter & Reference  & $R_{\textrm{1-to-2}}$ & Simulation (a) & $R_{\textrm{1-to-2}}$ & Simulation (b) & $R_{\textrm{1-to-2}}$ \\
  \hline
  $\beta$ & $3.2\times 10^{-6}$ & 4.7 & $5.8\times 10^{-6}$ & 4.1 & $6.4\times 10^{-6}$ & 4.1 \\
   $\gamma$ & $5.4\times 10^{-7}$ & 1.4 & $5.4\times 10^{-7}$ & 1.4 & $5.4\times 10^{-7}$& 1.4 \\
  $\eta$ & $1.3\times 10^{-5}$ & 4.7 & $2.3\times 10^{-5}$ & 4.3 & $2.5\times 10^{-5}$ & 4.0 \\
  $\alpha_1$ & $1.1\times 10^{-7}$ & 5.4 & $1.2\times 10^{-7}$ & 5.2 & $1.3\times 10^{-7}$ & 4.8 \\
  $\alpha_2$ & $4.5\times 10^{-8}$ & 2.2 & $5.0\times 10^{-8}$ & 2.2  & $5.0\times 10^{-8}$ & 2.6 \\
    $\mu_{\odot}$ & $2.1\times 10^{13}$ & 3.0 & $9.1\times 10^{13}$ & 2.1  & $1.1\times 10^{14}$ & 2.2 \\
  $J_{2\odot}$ & $1.1\times 10^{-9}$ & 1.7 & $1.1\times 10^{-9}$ & 2.1 & $1.1\times 10^{-9}$ & 2.2 \\
  $\zeta$ & $3.0\times 10^{-15}$ & 3.0 & $3.1\times 10^{-15}$ & 3.1 & $3.1\times 10^{-15}$ & 3.1 \\
  $GS_{\odot}$ & $6.8\times 10^{39}$ & 1.8 & $9.3\times 10^{39}$ & 1.4 & $9.4\times 10^{39}$ & 1.4 \\
  \hline
  $t_1$ & -- & -- & $8.6\times 10^{-6}$ & 2.0 & $5.7\times 10^{-5}$ & 3.3 \\
  $t_2$ & -- & -- & $6.5\times 10^{-6}$ & 2.0 & $4.4\times 10^{-5}$ & 3.2 \\
  $t_3$ & -- & -- & -- & -- & $1.4\times 10^{-4}$ & 2.6 \\
  \end{tabular}
  \end{ruledtabular}
\end{table*}
We can observe that the main benefits of an extended mission concern
the determination of $\beta$, $\eta$ and $\alpha_1$, whose accuracies
can be improved almost by a factor 5. This fact can be easily
understood considering that the effects of these parameters on the
planetary dynamics are more visible over long time scales, hence the
benefit of a longer mission. On the other hand, $\gamma$ and
$\mu_\odot$ would not benefit in any way from a duration longer than
one year. Regarding the torsion parameters, their knowledge could be
improved by a factor 2 by considering a 2-year long orbital phase.

\section{Discussion and conclusions}
\label{sec:5}

In this paper we have shown how to account for a possible
non-vanishing spacetime torsion in the framework of the MORE relativity experiment. In particular, we have described the implementation of the updated dynamical model within the ORBIT14 software, developed by the Celestial Mechanics
Group at the University of Pisa. The implementation has been done, in turn, by
parameterizing the torsion effects by means of three parameters,
$t_1$, $t_2$, $t_3$, to be estimated in a global LS fit in the same
way as the PN and related parameters, which characterize the
``classical'' MORE relativity experiment of BepiColombo.

We have shown that the torsion parameters can be estimated with MORE,
at least, at the level of some parts in $10^{-4}$, which is a
remarkable result in order to constrain torsion theories of
gravitation. Our analysis showed that the main limitations in
estimating the torsion are due to the correlation of the torsion parameters
with the gravitational mass of the Sun, $\mu_\odot$, and of $t_3$, in
particular, with the PN parameter $\beta$. This issue can be
overcome by adding an apriori constraint on $\mu_\odot$ and
$\beta$. The consequence is that, when all the three torsion
parameters are estimated in the global LS fit (i.e., simulation
scenario (b)), $\beta$ and $\mu_\odot$ cannot be determined at a
better level with respect to the apriori value. Thus, we can conclude
that the torsion parameters can be estimated at a significant level of
accuracy by the MORE relativity experiment, but this can be done, in
turn, by relaxing the requirements on the estimate of $\beta$ and
$\mu_\odot$.

On the other hand, any independent improvement in the knowledge of
$\beta$ and $\mu_\odot$ could be adopted as an apriori on these
parameters within our fit and it could allow furher improvement on
the determination of the torsion parameters. To quantify this point in
the case of $\beta$, we ran two illustrative simulations starting from
the scenario of simulation (b) but we tightened the apriori constraint
on $\beta$ up to $1\times 10^{-5}$ (case I) and $5\times 10^{-6}$
(case II). The results are shown in Table \ref{tab_res4}, where the
case of simulation (b) is compared with case I and case II for what
concerns the achievable accuracies on $\beta$, $t_1$, $t_2$, $t_3$.
\begin{table}
  \begin{ruledtabular}
    \caption{Comparison of the 1-$\sigma$ accuracies achievable for
      the torsion parameters, setting different apriori constraints on
      $\beta$: $3\times 10^{-5}$ (simulation (b)), $1\times 10^{-5}$
      (case I), $5\times 10^{-6}$ (case II).}\label{tab_res4}
  \begin{tabular}{cccc}
   Parameter & Simulation (b)  & Case I & Case II \\
  \hline
  $\beta$ & $2.6\times 10^{-5}$ & $9.8\times 10^{-6}$ &  $5.0\times 10^{-6}$ \\
  \hline
  $t_1$ & $1.9\times 10^{-4}$ & $1.4\times 10^{-4}$ & $1.4\times 10^{-5}$  \\
  $t_2$ & $1.4\times 10^{-4}$ & $1.1\times 10^{-4}$ & $1.0\times 10^{-5}$  \\
  $t_3$ & $3.6\times 10^{-4}$ & $3.4\times 10^{-4}$  & $3.4\times 10^{-4}$  \\
  \end{tabular}
  \end{ruledtabular}
\end{table}
Comparing the results of simulation (b) with respect to case I and II,
we can observe that a possible independent improvement on the
knowledge of $\beta$ up to $5\times 10^{-6}$ would not allow any
significant improvement in the determination of the torsion
parameters. This result suggests that an accuracy of some parts
in $10^{-4}$ is the ultimate level of accuracy by which the MORE
relativity experiment itself is capable to constrain the torsion
parameters.

%\subsection{Appendices}
%Appendices should be specified using the \m{appendix} command which
%specifies that all following sections create with the \m{section}
%commands are appendices. If there is only one appendix, then the
%\m{appendix*} command should be used instead.

\subsection{Acknowledgments}
The results of the research presented in this paper have been
performed within the scope of the contract "addendum
n. 2017-40-H.1-2020 all'accordo attuativo n. 2017-40-H" with the
Italian Space Agency (ASI).

% Create the reference section using Bib-TeX:

%\bibliographystyle{apsrev4-2}
\bibliography{bibliography}

\end{document}